# THE SMART PARKING MANAGEMENT SYSTEM

Amira. A. Elsonbaty[1] and Mahmoud Shams[2]

[1]Department of communication and electronics, Higher institute of engineering and technology, new Damietta, New Damietta, Egypt, 34517
[2]Department of Machine Learning and Information Retrieval, Faculty of Artificial Intelligence, Kafrelsheikh University, Kafrelsheikh, Egypt, 33511

*ABSTRACT*

*With growing, Car parking increases with the number of car users. With the increased use of smartphones and their applications, users prefer mobile phone-based solutions. This paper proposes the Smart Parking Management System (SPMS) that depends on Arduino parts, Android applications, and based on IoT. This gave the client the ability to check available parking spaces and reserve a parking spot. IR sensors are utilized to know if a car park space is allowed. Its area data are transmitted using the WI-FI module to the server and are recovered by the mobile application which offers many options attractively and with no cost to users and lets the user check reservation details. With IoT technology, the smart parking system can be connected wirelessly to easily track available locations.*

## 1. INTRODUCTION

The number of car client's increases was requested more parking spots, and with the growth of the internet of things causes smart urban areas to have picked up grind popularity. In this way, issues, for example, traffic blockage, constrained vehicle leaving offices, and street security are being tended to by IoT. So, several parking organization systems have been organized to decrease such traffic issues and improve the comfort of car users, it has combined smart mobiles, wireless algorithms, and mobile applications. The idea of the Internet of Things (IoT) started with things with Personal communication devices, which the devices could be tracked, controlled to use remote PCs connected with the internet [1]. The Internet of Things (IoT) equals "=" Physical devices, vehicles, structures, and different things implanted with hardware "+" Controller, Sensor, and Actuators "+" organize a network that lets these things to gather and exchange information (Internet) [2]. Sensors are deployed in smart systems, which in turn collect information from the device for processing and analysis .So, Sensors would be deployed in the parking area and through the mobile application for helping the user to know the freedom of parking places on a real-time basis with more efficiency, and less cost [3]. A smart parking system reduces the time to locate available places and reduces fuel consumption. The paper is organized as follows: First, it presents the concept of the smart parking system and its various functions, then its reviews previous research and studies on the implementation of smart parking. Then it describes the system implementation and operation and gives a conclusion of the smart parking application.

## 2. SMART PARKING SYSTEM

One of the most important problems facing large cities is congestion and parking . So, using Automated Parking System Management is an efficient technique using the Internet of Things to manage the garage [4]. Smart parking is an electronic tool that enables the user to find vacant parking spaces through information technology and by using appropriate sensors [5]. Among the most used types in smart parking, systems are data routing systems, smart payment systems, and electronic car parks. These types require disclosure of whether parking spaces are vacant or not. With the user registration in the system, a unique identifier is created for him, and with the booking, it has the booking details, and via their smartphones, the entire time, exit time, and money are calculated. The System building consists of, the lowest level, including the functions of sensing, data transmission is created during a middle level, and upper-level deals with the storage and processing information, and user interfaces [6].

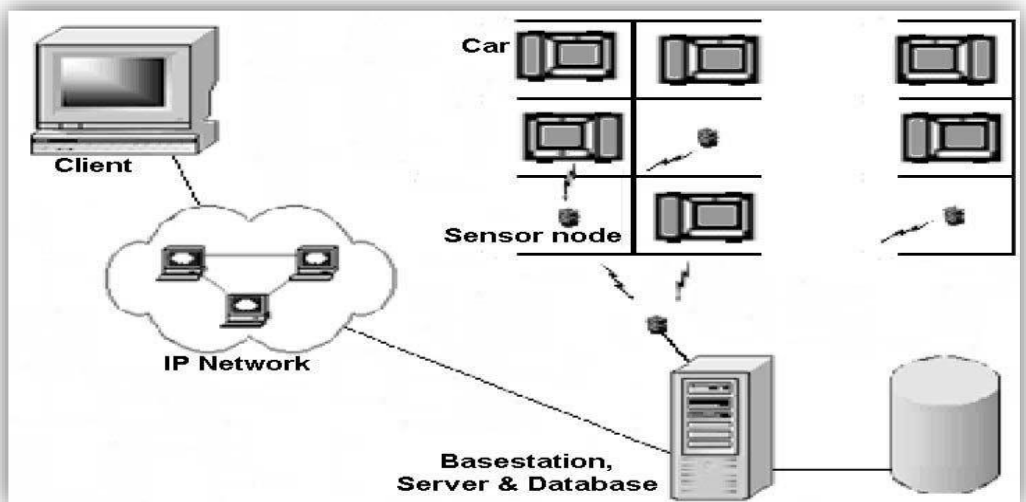

Figure 1. Smart parking system building

## 3. LITERATURE SERVEY

In this section, some related works of smart parking are presented. The System [7] was used sensors, technologies, and interfaces to collect and display information in real-time, which required expensive private infrastructure. The smart parking system [8] delivered user information and accessibility of parking slots through the VMS on the internet. It was classified into off-road and on-road. Google map application, ultrasonic sensors, and cloud-stored data were used in Smart parking [9], the Android application map forward data of the empty place of the user. Each slot had an LED for finding out the parking space and booking. The Infrared sensor in [10] was implemented to find out a free place and open the entry and exit gate. RFID tag issued to approve an individual's entry to the parking place using the mobile application the ACO algorithm was presented in [11] to calculate the shortest path between the user, and the available space by connecting to the cloud. The System [12] using Arduino and Raspberry PI to detect the free slots by using an internet server, and GPS for booking. The system [13] was proposed which uses Infrared sensors. Verification is accomplished using an RFID tag and ZigBee for communication. The android application in [14] would have customer detail include area, state, car number, the user enters and exit time, and selecting a parking location gets data about the available vacant parking slot, and MYSQL database stored user details. The system [14] was employed a camera to capture the car number plate and convert the image to see whether the car is a certified used car or not. The system [15] used Vehicle to Infrastructure (V2I) to communicate with the driver sending the parking request providing, the user information status of conforming reservation, and Infrastructure to Vehicle (I2V) communication was employed for reserve parking place application and shows the direction. For safety purposes, QR code was employed, webcam used to scan the code, and authorized to indicate the car parking zone direction. A Privacy-Preserving Pay-by-Phone parking system [16] was proposed, which book by pay by phone method. Mobile application using credit card payment method is implemented. The new user can register, and the new user contacts the system, server, and gets a new e-coins. Each e-coin having a parking duration time of the slot. Parking officer queries of on-board devices by performing an RFID query. The system [17] delivers parking facilities in real-time and users are ready to book places and make payments before incoming at the car park's lot. The system in [18] proposed a communication system and a database using the cloud. An ultrasonic sensor in [18] placed on the ground connected via Ethernet. The system [19] used wireless communication to order places of the nearest parking spaces through the GPS. The system transmits the availability of spaces every 2 mins. If all parking spaces are unavailable, no actions are measured; within the opposite case, any user is ready to order a locality within 2 km of their location. A system was presented in [20] calculates the optimal car parking, neighborhood for the user-supported space of trajectory and time. The system did not have the booking service and was subject to the availability of the space at that time.

## 4. THEORY OF THE PROJECT

### 4.1. The Problem Definition

People usually travel around within the parking regions trying to find an appropriate place to park in, to solve this problem, the automated car parking system has created. Assistive technology is needed, which may provide parking information for registered customers using smartphones and their applications. Users can obtain the service by registering, and in case of booking, the destination and the estimated time of arrival are determined, and the booking details are sent to the user.

### 4.2. Aim of the project

The smart car parking system is an integrated system to recognize the nearest available parking zone. So, the main purpose of the system is to provide a solution to the parking problem, to reduce the time to search for parking lots, and to eliminate unnecessary travel for vehicles

### 4.3. How Smart Parking Works

Smart parking suggests an IoT-based system that sends data to free and busy parking places via net/mobile applications. The IoT-network includes sensors and microcontrollers, which are found in each parking place. We implemented an enclosed smart parking project (SPMS), that using the Internet of Things and IR sensors, where available parking places can be displayed in a web application, then the user receives a live update about the availability of all parking places and chooses the best one. Smart parking IoT implementation is usually divided into the following parts:

| 1 | Collection | The collection depends on parking sensors to collect real-time parking. The parking systems may use sensors like Infrared, and Ultrasonic Sensorsdetect whether a parking slot is empty or not [21]. Also, an ESP8266 Wi-Fi chip comprises ofthe TCP / IP protocol, that licenses any microcontroller to contact a Wi-Fi network. |
|---|---|---|
| 2 | Processing | The processing unit acts as interference between the sensors and the cloud [22]. It includes an Arduino which is a processor on-chip. All the sensors are wirelessly connected to the processing unit, and data collected from various sensors are sent to it through the esp8266 chip. |
| 3 | Deployment | It deals with communication methods. Message Queue Telemetry Transport Protocol (MQTT) is a publish-subscribe based messaging protocol that is used on top of the TCP/IP protocol [23]. |
| 4 | Services | It can be made available to users once they finish storing data and monitoring information. |
| 5 | Connection | Interested in the Internet of Things layer that deals with the database of parked cars through a shared server. The cloud stores data for available parking lots, user sites, profiles, etc. [24].It keeps a track of each user connected to the system andstoresa backup of the information stored in the cloud. |
| 6 | Mobile application | It is the interface application between humans and the system. |

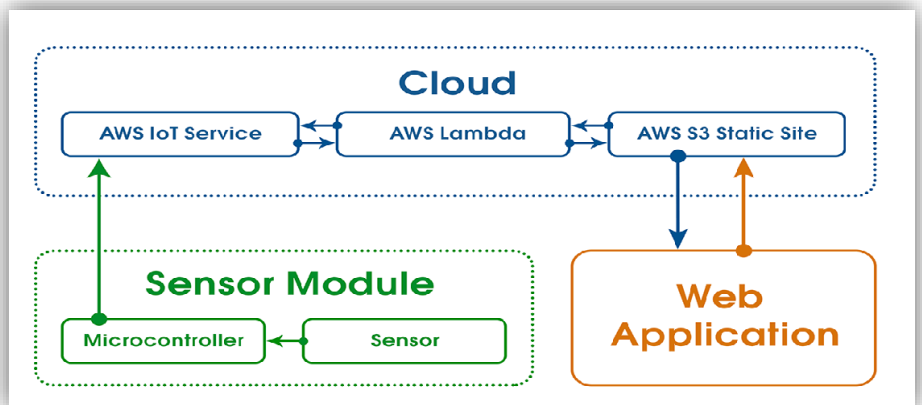

Figure 2. Smart Parking System.

## 5. IMPLEMENTATION & WORKING

### 5.1. The Proposed System

Finding a place to park cars involves three-stage. First, the parking area which has Arduino devices along with the sensors to interact between the user and the parking area. The second stage contains the cloud services which act as an intermediary between the user and the parking area. The third stage is the user side. The user gets a notification of the availability via mobile applications. For each parking region, Arduino sensors are positioned, and the sensors detect the number of parking slots, the number of free, and booked slots. WIFI module is used for communication between the mobile app and sensors.

International Journal of Computer Science & Information Technology (IJCSIT) Vol 12, No 4, August 2020

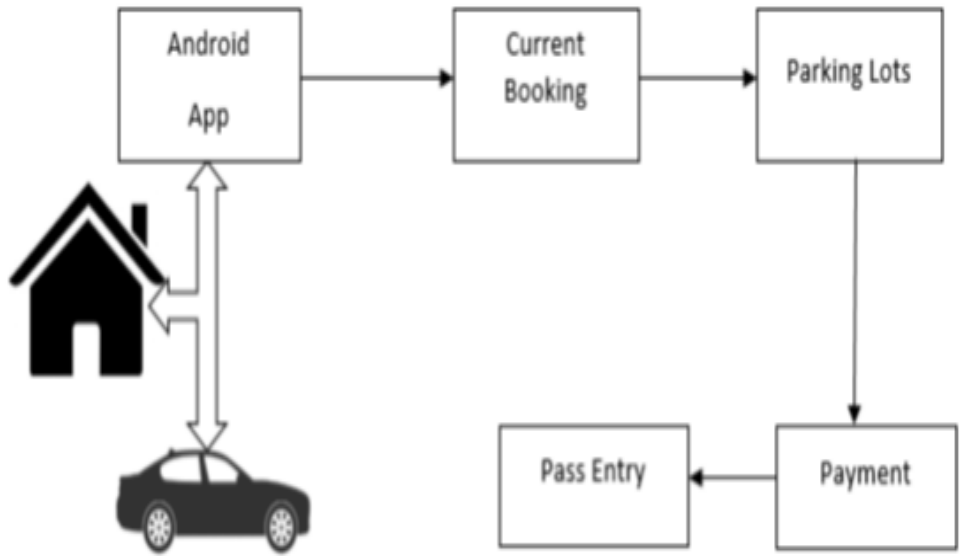

Figure3. The architecture of booking for parking.

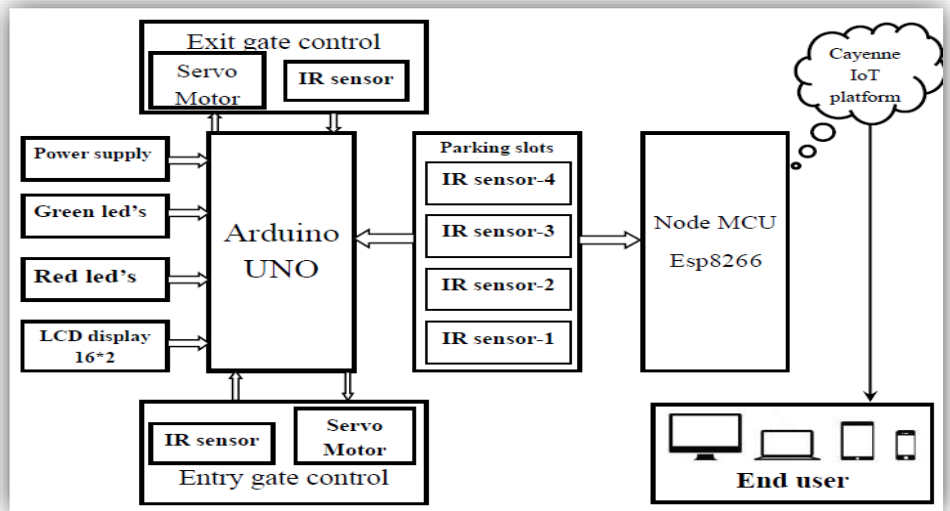

Figure 4. The smart parking implementation

### 5.2. Components of Proposed System

The proposed system works through a set of commands within the Arduino and it needs hardware components to work suitably.

### 5.2.1 Hardware Components& Circuits

1. **Arduino** is a project created by the largest technical community of engineers and developers to develop interactive control projects using various types of electronic boards programmed with free programming language [25].

International Journal of Computer Science & Information Technology (IJCSIT) Vol 12, No 4, August 2020

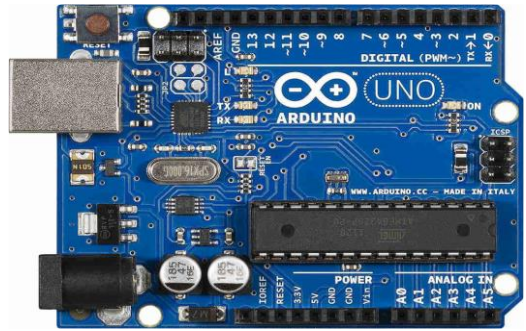

Figure 6, Arduino UNO Board.

2. **GSM Module** is a circuit which is used to set up communication between mobile phones and microcontroller. It is used to send SMS, MMS, and voice messages through a mobile network [26].

3. **Node MCU** is an open-source Lua built firmware and development board targeted for IoT based applications, that relies on the ESP8266 Wi-Fi, and the ESP-12 module [27].

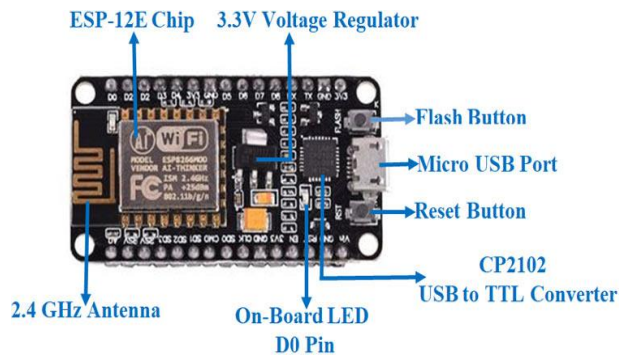

Figure 7, NodeMCU ESP8266 development board

4. 16×2 LCD Display with I2C is an electronic display module that produces a visible image that can display up to 32 characters on a single screen [28]. LCD Display uses many numbers of Pins of Arduino for connecting Inter-integrated Circuit (I2C). It decodes the data received from the I2C Bus into Parallel data that is required for the LCD Display.

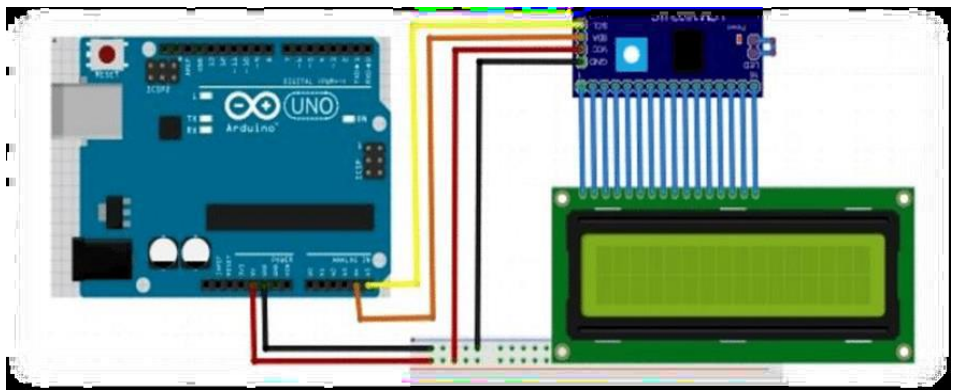

Figure 8. 16×2 LCD Display with I2C

60

International Journal of Computer Science & Information Technology (IJCSIT) Vol 12, No 4, August 2020

5. TCRT5000 Circuit: An Infrared (obstacle sensor) uses to detect the presence of the object or any other reflective surface in front. Its package has a Photodiode that uses to generate an IR signal and a Phototransistor which can be used to read the IR signal that is reflected [29]. The obstacle detected if the reading of the IR sensor is "0".

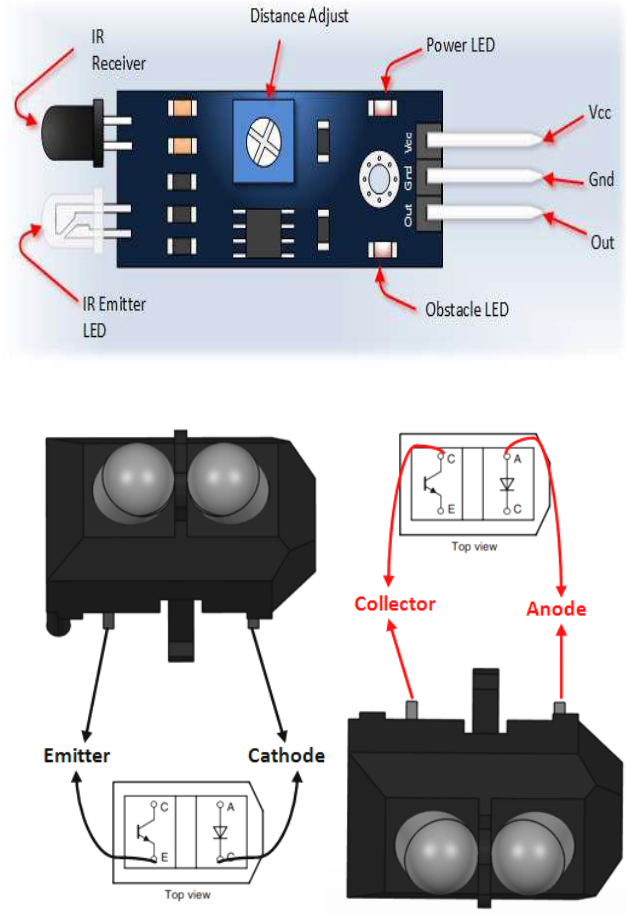

Figure 9. TCRT5000 Circuit

6. A servo motor is a motor with a gearbox and a Shaft transmission that gives motion greater torque and greater precision [30]. When the engine is pulsed at a certain time, the engine rotates the angle according to that time.

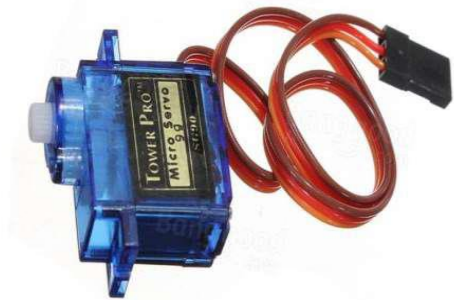

Figure 10. A servomotor

7. Piezoelectric Sensor converts physical parameters, for example, acceleration, strain, or pressure into an electrical charge which can then be measured [31].

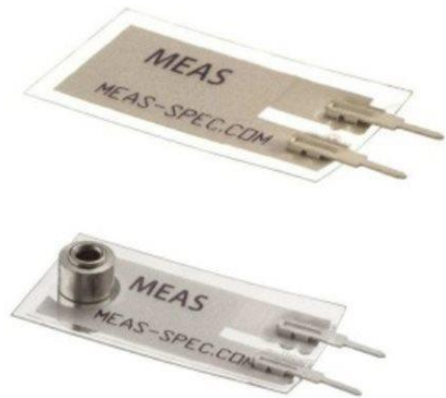

Figure 11. Piezoelectric Sensor

### 5.2.2  Second: Software Components & Circuits

1. Arduino IDE supports the languages C and C++ using distinct guidelines of code architecture, which stores a software library from the wiring project, which runs common input and output procedures.

2. MS SQL Server is a client-server architecture that accepts, processes, and replies to the client request with processed data.

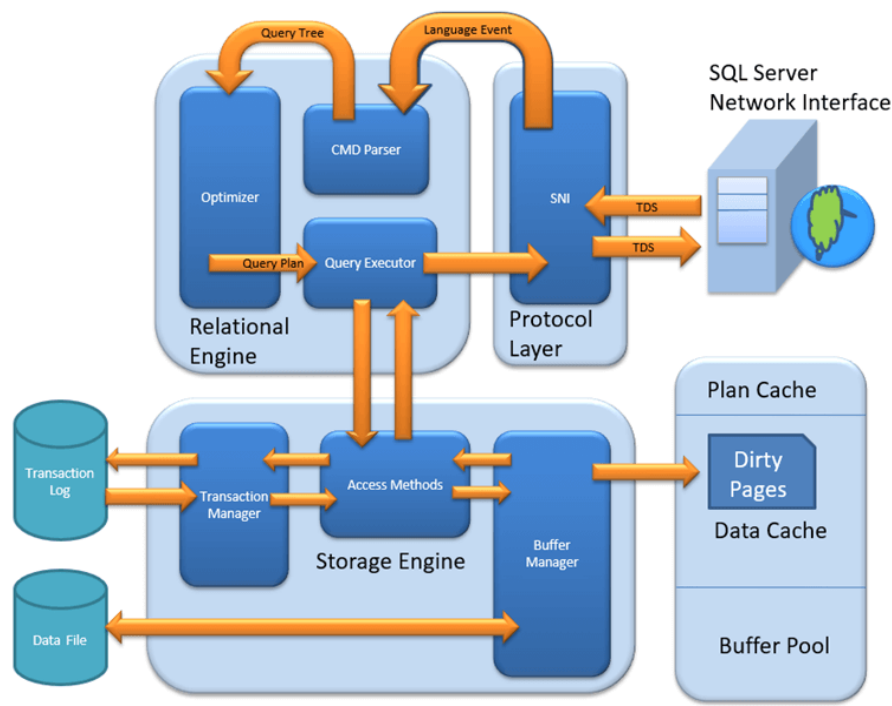

Figure. 12. SQL Server Architecture Diagram

## 6. RESULTS

### 6.1. Hardware Implementation

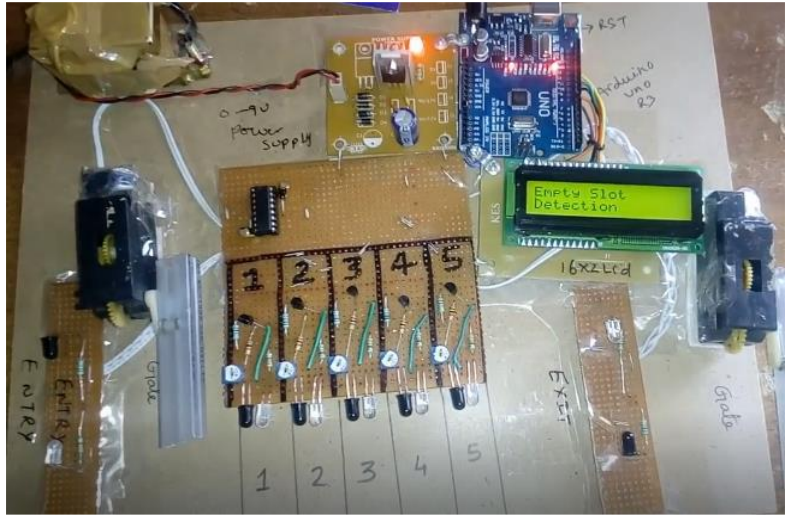

Figure. 13, smart parking implementation

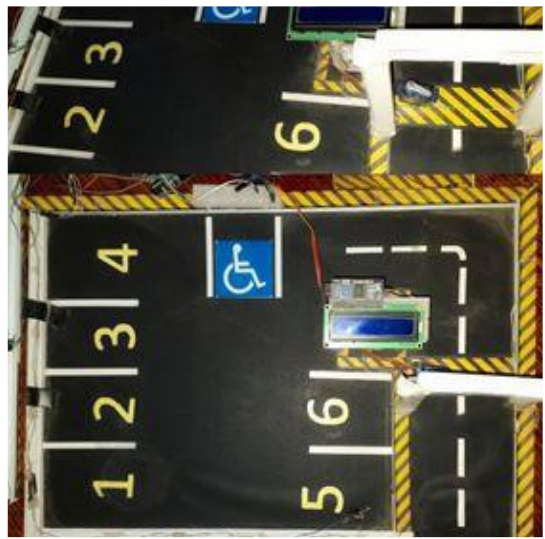

Figure. 14, smart parking design

International Journal of Computer Science & Information Technology (IJCSIT) Vol 12, No 4, August 2020

## 6.2. Smart Parking Database

The tables and the relations of the smart parking database are shown in figure 16

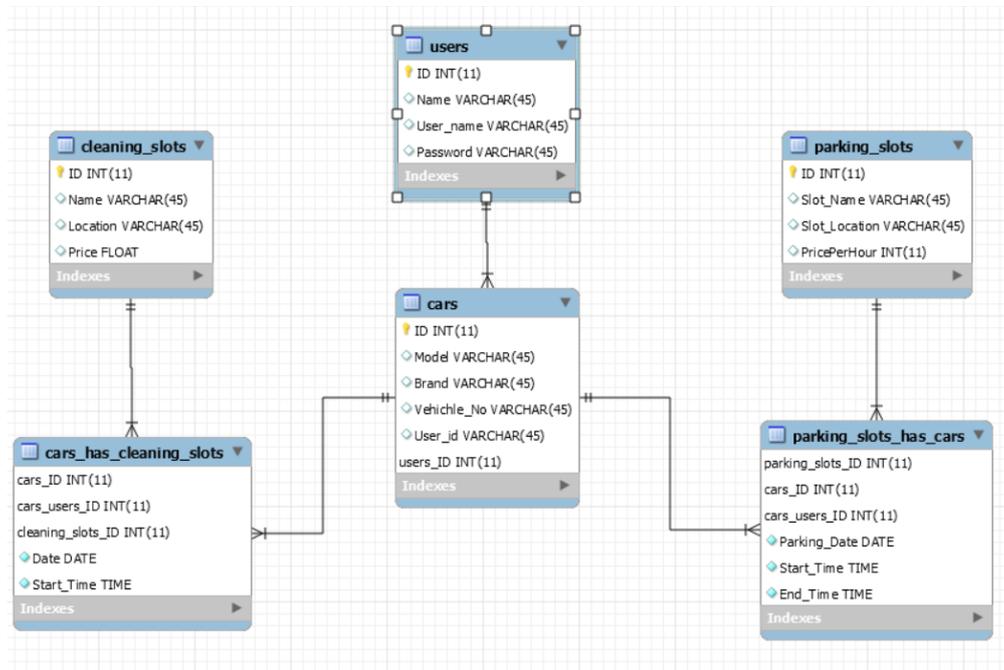

Figure. 15. Tables & Relations of smart parking database

6.3. **Android Parking App** is developed using the Android Studio application platform. Figure 17 (A- B- C- D- E- F- G- H) displays the android mobile application pages.

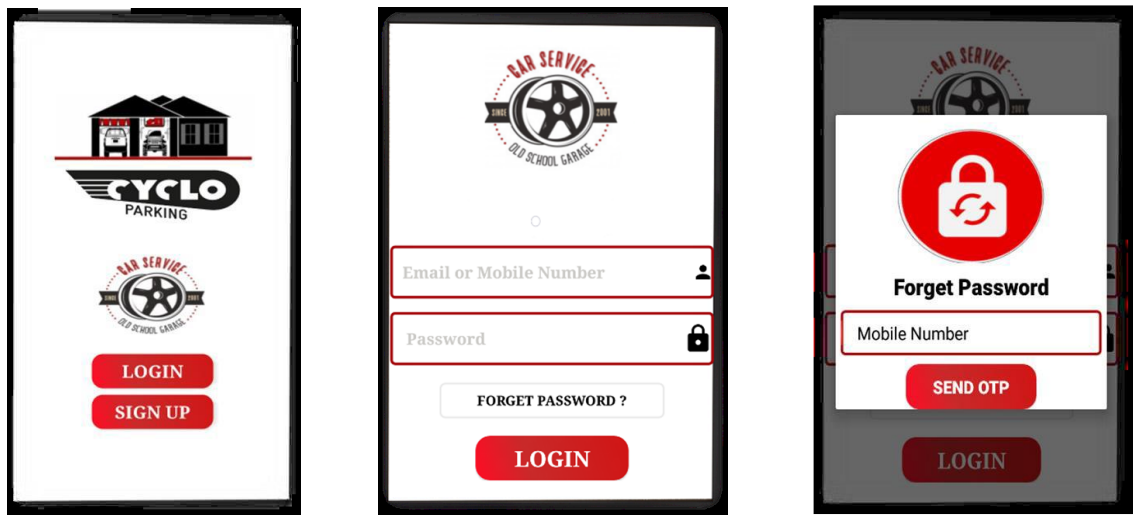

A. Splash screen of application   B. Login page   C. Reset forget password

International Journal of Computer Science & Information Technology (IJCSIT) Vol 12, No 4, August 2020

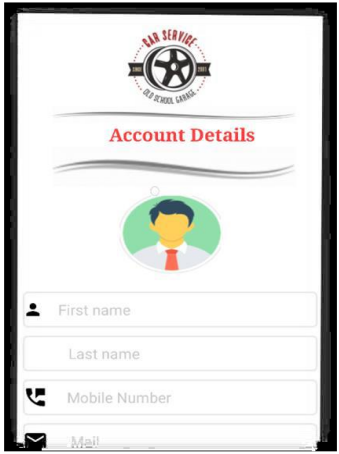
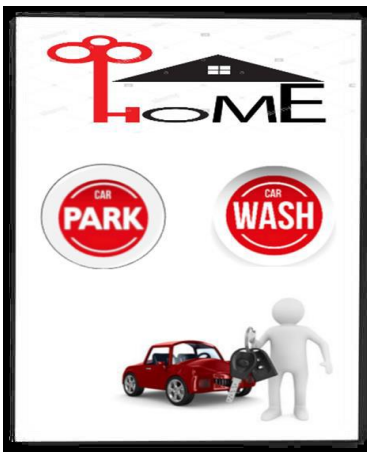
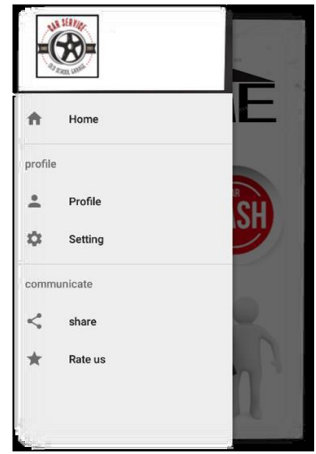

D. Registration page      E.-Select service      F. Extra service

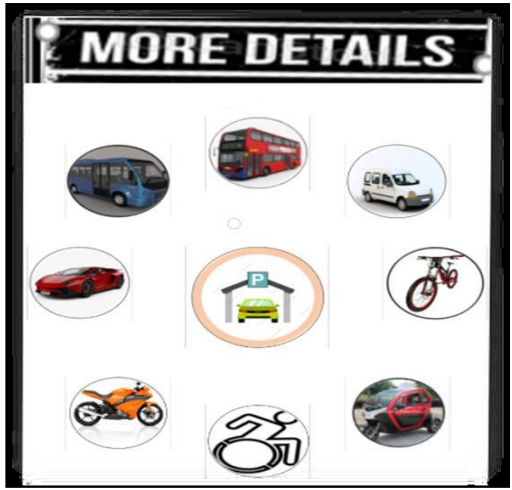
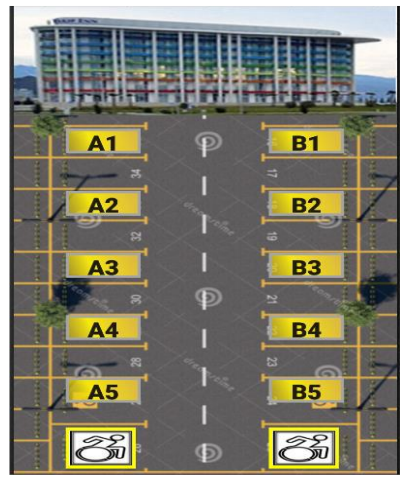

G. Searching page for parking      H. View slots of parking

## 7. CONCLUSION

The services provided by smart parking have become the essence of building smart cities. This paper focused on implementing an integrated solution for smart parking. The proposed system has several advantages, including detecting parking spaces using the Internet of Things and calculating the time of entry and exit and calculating the expected cost. An attractive and effective application was designed for Android mobile phones. The system benefits from avoiding wasting time and reducing pollution and fuel consumption. Users can book a car park for 24 hours.

## REFERENCES

1. Abhirup Khanna, Rishi Anand, "IoT based Smart Parking System", Proc., In 2016 International Conference on Internet of Things and Applications (IOTA), 22 Jan - 24 Jan 2016.
2. Anusha, Arshitha M, S, Anushri, Geetanjali Bishtannavar "Review Paper on Smart Parking System," International Journal of Engineering Research & Technology (IJERT), ISSN: 2278-0181, Volume 7, Issue 08, Special Issue – 2019.
3. S. Senthil, M. Suguna, J. Cynthia, "Mapping the Vegetation Soil and Water Region Analysis of Tuticorin District Using Landsat Images", IJIEST ISSN (2455-8494), Vol.03, No. 01, Jan 2018.

4. Juhi Seth, Pola Ashritha, R Namith, "Smart Parking System using IoT ElakyaR", International Journal of Engineering and Advanced Technology (IJEAT), ISSN: 2249 – 8958, Volume-9 Issue-1, October 2019.
5. Mimbela, L.Y. and L.A. Klein, "A summary of vehicle detection and surveillance technologies used in intelligent transportation systems", New Mexico State University, Tech. The report, 2007.
6. M. Y. I. Idris, Y. Y. Leon, E. M. Tamil, N. M. Noor, and Z. Razak, "Car parking system: A review of the smart parking system and its technology," Information Technology Journal, pp. 101-113.], 2009.
7. Paidi. V; Fleyeh, H.; Hakansson, J.; Nyberg, R.G.," Smart Parking Sensors, Technologies and Applications for Open Parking Lots: A Review", IET Intel. Transport Syst, 12, 735–741, 2018.
8. Amir O. Kotb, Yao-Chunsheng, and Yi Huang "Smart parking Guidance, Monitoring and Reservation: A Review," IEEE-ITSM, pp.6-16. Apr-2017.
9. Supriya Shinde, AnkitaM Patial, pSusmedha Chavan, Sayali Deshmukh, and Subodh Ingleshwar, "IOT Based Parking System Using Google", Proc., of. I-SMAC,2017, pp.634-636, 2017.
10. Hemant Chaudhary, PrateekBansal., B. Valarmathi," Advanced CAR Parking System using Arduino", Proc., of. ICACCSS, 2017.
11. Wang, M.; Dong, H.; Li, X.; Song, L.; Pang, D. A Novel Parking System Designed for G. Searching page for parking H. View slots of parking Smart Cities. Proc., in 2017 Chinese Automation Congress (CAC), Jinan, China, pp. 3429–3434, 20–22 October 2017.
12. Nastaran Reza NazarZadeh, Jennifer C. Dela," Smart urban parking deducting system", Proc., of. ICSCE, 2016, pp-370-373,2016.
13. PavanKumarJogada and VinayakWarad, "Effective Car Parking Reservation System Based on Internet of things Technologies ", Proc., of. BIJSESC, Vol. 6, pp.140-142, 2016.
14. Yashomati R. Dhumal, Harshala A. Waghmare, Aishwarya S. Tole, Swati R. Shilimkar," Android Based Smart Car Parking System" Proc., of. IJREEIE, Vol. 5, Issue 3, pp-1371-74, mar-2016.
15. Faiz Ibrahim Shaikh, Pratik NirnayJadhav, Saideep Pradeep Bandarakar" Smart Parking System based on embedded system and sensor Network" IJCA, vol.140. pp.45-51. Apr-2016.
16. RicardGarra, Santi Martinez, and Francesc Seb"e" A Privacy-Preserving Pay-by-phone Parking system" IEEE-TVT, pp.1-10, Dec-2016.
17. Khanna, A.; Anand, R.," IoT based Smart Parking System", proc., in 2016 International Conference on Internet of Things and Applications (IOTA), Pune, India, 22–24 January 2016; pp. 266–270.
18. Karthi, M.; Preethi, H. Smart Parking with Reservation in Cloud-based environment. In Proceedings of the 2016 IEEE International Conference on Cloud Computing in Emerging Markets, Bangalore, India, 19–21 October 2016; pp. 164–167.
19. Orrie, O.; Silva, B.; Hancke, G.P. "A Wireless Smart Parking System", prco., in 41st Annual Conference of the IEEE Industrial Electronics Society (IECON), Yokohama, Japan, pp. 4110–4114, 9–12 November 2015.
20. Hsu, C.W.; Shih, M.H.; Huang, H.Y.; Shiue, Y.C.; Huang, S.C., "Verification of Smart Guiding System to Search for Parking Space via DSRC Communication", Proc., in 12th International Conference on ITS Telecommunications, Taipei, Taiwan, pp. 77–81, 5–8 November 2012.
21. Revathi, G., & Dhulipala," Smart parking systems and sensors: A survey", proc., in 2012 International Conference on Computing, Communication, and Applications, 2012.
22. Abhirup Khanna, Rishi Anand," IoT based Smart Parking System", proc., in International Conference on Internet of Things and Applications (IOTA) Maharashtra Institute of Technology, Pune, India 22 Jan - 24 Jan 2016.
23. https://en.wikipedia.org/wiki/MQTT, 18-7-2020.
24. Thusoo, A.; Sarma, J.S.; Jain, N.; Shao, Z.; Chakka, P.; Zhang, N.; Antony, S.; Liu, H.; Murthy, R. HIVE-A,"petabyte-scale data warehouse using Hadoop", proc., In 2010 IEEE 26th International Conference on Data Engineering (ICDE 2010), 2010.
25. https://www.arduino.cc, 18-7-2020.
26. ElakyaR, Juhi Seth, Pola Ashritha, R Namith," Smart Parking System using IoT ", International Journal of Engineering and Advanced Technology (IJEAT) ISSN: 2249 – 8958, Volume-9 Issue-1, October 2019.
27. https://store.fut-electronics.com,18-7-2020.
28. https://www.instructables.com, 18-7-2020.
29. https://components101.com/sensors/tcrt5000-ir-sensor-pinout-datasheet. 18-7-2020.
30. https://engineering.eckovation.com, 18-7-2020.
31. https://www.variohm.com, 18-7-2020.